\documentclass[twocolumn,aps,pre,showpacs,epsfig,amsmath,amssymb,eqsecnum]{revtex4}
\usepackage{multirow}
\usepackage{pdflscape}
\usepackage{bbm}
\usepackage{mathrsfs}
\usepackage{rotating}
\usepackage{amsfonts}
\usepackage{psfrag}
\usepackage{color}
\usepackage{graphicx}
\def\newblock{\hskip .11em plus .33em minus .07em}

\newcommand{\be}{\begin{equation}}
\newcommand{\ee}{\end{equation}}
\newcommand{\bea}{\begin{eqnarray}}
\newcommand{\eea}{\end{eqnarray}}
\begin{document}
\title{Renormalized Surface Charge Density for a Strongly Charged Plate
 in  Asymmetric Electrolytes: Asymptotic Exact Results in Poisson Boltzmann Theory}
\author{Ming Han}
\address{Zhiyuan College,
Shanghai Jiao Tong University,
Shanghai, 200240 China}
\email{hanmingcr@163.com}
\author{Xiangjun Xing}
\address{Institute of Natural Sciences and Department of Physics,
Shanghai Jiao Tong University,
Shanghai, 200240 China}
\email{xxing@sjtu.edu.cn}

\date{\today} 
\begin{abstract}
The Poisson-Boltzmann equation for a strongly charged plate inside a generic charge-asymmetric electrolyte is solved using the method of asymptotic matching.  Both near field and far field asymptotic behaviors of the potential are systematically analyzed.  Using these expansions,  the renormalized surface charge density is obtained as an asymptotic series in terms of the bare surface charge density.
\end{abstract}

\maketitle
\section{Introduction}
\label{sec:intro}

A charged object inside an electrolyte attracts ions of opposite charge and repels ions of like charge.  The total electrostatic potential, due to both the external charges and the electrolyte, is described by the Poisson-Boltzmann (PB) equation~\cite{Andelman-PBequation} at the mean field level.  Study of PB is a significant problem in both physical and biological sciences.  Even though it has been clear in recent years that the Poisson-Boltzmann theory is inadequate for strongly charged systems \cite{Levin-charge-review,RevModPhys.74.329, Boroudjerdi2005129}, it remains a benchmark theory in interpreting experimental data on charged macromolecules, particles, and surfaces.  Despite of its importance, our analytic understanding of PB is very limited, largely due to its nonlinear nature.  Even for the simplest problem of one charged plate, exact solutions are known only for cases of $1:-1$ and $2:-1$ electrolytes \cite{Tracy-Widom-Physica-1997}.  

Consider an electrolyte with concentration $c_+$ of $m$ valence positive ions and concentration $c_-$ of $-n$ valence negative ions.  It is {\em asymmetric} if $m \neq n$. Inside this electrolyte, and at mean field level, the average electrostatic potential
near a charged plate satisfies the celebrated Poisson-Boltzmann equation (PB) \footnote{We will use SI units in this work. }:
\be
 - \epsilon  \phi''(z) + c_-  n q \,e^{n \beta q \phi(z)}
 - c_+ m q \, e^{ - m \beta q \phi(z)} = 0,
\label{PB-asymmetric}
\ee
where $z$ is the coordinate perpendicular to the plate, $\beta =  1/ k_B T$, and $q= 1.6\times 10^{-19}C$ the charge of an electron.  Because the dielectric constant of the plate is much smaller than that of the solvent, the standard electrostatic boundary condition reduces to that of Neumann:
\be
- \epsilon \phi'(z_0) = \sigma,
\label{Neumann-0}
\ee
where $z_0$ is the location of the plate, and $\epsilon$ dielectric constant of the solvent,  and $\sigma$ the surface charge density on the plate.  It will become clear below why we choose to locate the plate not at the origin of coordinate system. Note that the overall charge neutrality of the electrolyte imposes the constraint $m c_+ = n c_-$.

Far away from the charged plate, the potential must decay to its bulk value, which is chosen to be zero.  Therefore in the far field the PB can be linearized:
\be
- \Delta \phi + \kappa^2 \phi = 0,
\label{linear-PB}
\ee
where
\be
\kappa^2 = \ell_{DB}^{-2} = \frac{\beta q^2}{\epsilon}  (c_+ m^2 + c_- n^2),
\ee
with $\ell_{DB}$ the {\em Debye length}.   It is the characteristic length scale over which the electrostatic potential decays in the bulk.   The far field potential is therefore:
\be
\phi(z) \sim \frac{\sigma_R }{\epsilon\kappa}\, e^{-\kappa \delta z}, \quad
z \rightarrow \infty,
\label{far-field-general}
\ee
with $\delta z = z - z_0$ the distance from the plate to the field point.   $\sigma_R$ is called the {\em renormalized surface charge density}.  It acquires this name because Eq.~(\ref{far-field-general}) is the solution to linear PB Eq.~(\ref{linear-PB}) with {\em renormalized} boundary condition:
\be
- \epsilon \phi'(z_0) = \sigma_R.
\label{Neumann-RG-0}
\ee
Physically, $\sigma_R$ is the effective surface charge density we infer from the far field asymptotics if we use the linear PB equation.  Evidently, the value of $\sigma_R$ can only be determined by comparing the linear solution Eq.~(\ref{far-field-general}) with the solution to the full nonlinear PB Eq.~(\ref{PB-asymmetric}).   Calculation of the renormalized charge density $\sigma_R$ as a function of the {\em bare surface charge density} $\sigma$ is the problem that we shall tackle in this work.

It is convenient to rescale coordinate by the Debye length $\ell_{DB}$ and to
rewrite PB in terms of the dimensionless potential $\Psi = \beta q \phi$:
\be
-  \Psi''(z) + \frac{1}{m+n} \left(
e^{n\Psi} - e^{-m\Psi} \right) = 0.
\label{PB-Psi}
\ee
This equation can be integrated once, with the constant of integration determined by the boundary condition at infinity: $\Psi(\infty) = \Psi'(\infty)=0$:
\be
- \frac {\Psi'(z)^2 }{2}+\frac 1 {m+n} \left( \frac {e^{n\Psi}}{n} + \frac {e^{-m\Psi}}{m}- \frac 1{n}- \frac 1{m} \right)=0.
\label{first-integral}
\ee
The Neumann boundary condition Eq.~(\ref{Neumann-0}) reduces to
\be
- \Psi'(z_0)=   \frac{\beta q \sigma}{\epsilon \kappa}
\equiv \eta,
\label{Neumann-eta}
\ee
where $\eta$ is the {\em dimensionless bare surface charge density}.
The renormalized boundary condition Eq.~(\ref{Neumann-RG-0}) then reduces to
\be
-  \Psi'(z_0) = \eta_R,
\label{Neumann-RG}
\ee
where
\be
 \eta_R = \frac{\beta q}{\epsilon \kappa}\sigma_R,
 \label{eta-sigma-R}
\ee
is the dimensionless version of the renormalized charge density.  One of the main purpose of this work is to calculate the functions $\eta_R(\eta)$ for general cases of asymmetric electrolytes.  As we shall find, these functions only depends on the type of electrolyte $(m:-n)$, but are independent of other parameters such as temperature and dielectric constant of the solvent.   Using Eq.~(\ref{Neumann-eta}) and Eq.~(\ref{eta-sigma-R}), we can represent the renormalized charge density $\sigma_R$ in terms of the function $\eta_R(\eta)$ as
\be
\sigma_R = \frac{\epsilon \kappa} {\beta q} \, \eta_R(\eta)
=   \frac{\epsilon \kappa} {\beta q} \,
\eta_R \left(\frac {\beta q \sigma}{\epsilon \kappa}\right),
\ee
which is sufficient to determine the leading order behavior of far field asymptotics Eq.~(\ref{far-field-general}).

For $(m,n) = (1,1)$, and $(2,1)$ the exact solutions to PB Eq.~(\ref{PB-asymmetric}) are known.  From these, it is seen $\eta_R$ approaches a finite constant $\eta_R(\pm \infty)$ as the bare charge density approaches $\pm \infty$.
\be
\begin{array}{ll}
\eta_R(\pm \infty) = \pm 4 \quad & 1:-1,  \vspace{3mm}\\
\eta_R(\infty) = 6 \quad  & 2:-1 \,\, \mbox{positive plate}, \vspace{3mm}\\
\eta_R(- \infty) = -\frac{6}{2+ \sqrt{3}} \quad & 2:-1\,\, \mbox{negative plate}.
\end{array}
\ee
Therefore in PB theory, the potential in the bulk remains finite even when the bare charge density goes to infinity.   The full functions $\eta_R (\eta)$ are also known.  For details, see Table \ref{Table:eta-R-expansion}, also see reference \cite{Xing:2011fk}.

That $\eta_R(\pm \infty)$ saturates is a general property of nonlinear PB theory, independent of the shape of the charged objects. \footnote{For a proof for plate geometry, see reference \cite{Tellez:2003kx}.  The independence of geometry follows from the fact that saturation is a short scale property. } Already exhibited by the exact solution to one dimensional PBE within symmetric electrolyte which has been known for a long time, the significance of this property has only been appreciated recently \cite{Xing:2011fk,Trizac:2003kx,Trizac:2002uq,Belloni198543,Palberg:1995uq,Belloni-charge-RG}. As one important consequence, it implies that the interaction between two charged colloids saturates when their charges scale up, with their separation fixed.  This was seen in numerical solution of PB equation \cite{charge-RG-Alexander-JCP-1984} of a charged sphere, where the terminology {\em charge renormalization} was first introduced.  


In this work, we shall solve PB for one plate problem in a generic $m:-n$ asymmetric electrolyte.  We shall first develop asymptotic expansions of the electrostatic potential both in near field, for a plate with {\em infinite} bare surface charge density, located at the origin of axis $z = 0$ (Sec.~\ref{sec:near}).  From these expansions, it becomes clear that the potential does remain finite even if the bare surface charge density is infinity.  We shall then develop asymptotic expansions in the far field, of which Eq.~(\ref{far-field-general}) is the leading order (Sec. \ref{sec:far}).   In Sec.~\ref{sec:matching}, we shall match the far field expansion with the near field expansion, and determines the renormalized surface charge density $\eta_R$ for the infinitely charged plate.   Finally in Sec.~\ref{sec:highly-charged}, we exploiting the properties of near field expansion to obtain an asymptotic expansion of the renormalized charge density $\eta_R$ in terms of the bare density $\eta$.  It is important to point out that our method can be immediately generalized to the cases of  electrolyte mixtures.

Before starting to solve Eq.~(\ref{first-integral}), let us discuss two symmetries of this equation.   Let us use $\Psi_{\pm}^{m,-n}$ to denote a solution to Eq.~(\ref{first-integral}) inside $m:-n$ electrolyte, generated by a positively/negatively charged plate.  Eq.~(\ref{first-integral}) is a first order ODE, and therefore contains only one parameter in its general solution, which can be chosen to be the location of the plate.  Therefore the solution to Eq.~(\ref{first-integral}) is unique up to the translation of the plate.  Firstly, it is easy to check that $-\Psi_{\pm}^{m,-n}$ is a solution to PB inside an $n:-m$ electrolyte, generated   by a {\em negatively/positively} charged plate.  We therefore find the following relation:
\be
\Psi_-^{n,-m} (z) = -\Psi_+^{m,-n} (z).
\ee
From this we obtain the following relation:
\be
\eta_R^{m:-n}(\eta) = - \eta_R^{n:-m}(-\eta).
\label{eta_pm}
\ee
As a consequence, we only need to solve Eq.~(\ref{first-integral}) for all cases with $m \geq n$.

Secondly, if $m$ and $n$ have a common integer factor $p$, such that $m = p \, \tilde{m}, n = p \, \tilde{n}$, then we can define $\tilde{\Psi} = p \Psi$, which satisfies
\be
- \tilde{\Psi}'' + \frac{1}{ \tilde{m} + \tilde{n}}
\left( e^{\tilde{n} \tilde{\Psi}} - e^{- \tilde{m} \tilde{\Psi}}
\right) = 0.
\ee
Hence
\be
\Psi_{\pm}^{m:-n}(z) = p^{-1} \Psi_{\pm}^{\tilde{m}:-\tilde{n}}(z).
\ee
From this, we obtain the following relation between renormalized charge densities for the renormalized surface charge density:
\bea
\eta_R^{m:-n}(\eta) &=& p^{-1} \eta_R^{\tilde{m}:-\tilde{n}}(p \,\eta),\label{p}\\
\eta_R^{m:-n}(\pm \infty) &=& p^{-1} \eta_R^{\tilde{m}:-\tilde{n}}(\pm \infty).
\eea
As a consequence, we only need to solve Eq.~(\ref{first-integral}) for all $m, n$ that are relatively prime.


\section{Near Field Asymptotics}
\label{sec:near}
Sufficiently close to a strongly charged plate, the co-ions which carry like charges as the plate are strongly repelled so that their density is negligibly small.  Consequently, we may ignore the corresponding term in the Poisson-Boltzmann equation.   For a positively charged plate, the equation Eq.~(\ref{PB-Psi}) (with positively charged ions ignored) reduces to
\be
- \Psi_{0+}''(z) + \frac{1}{m+n} e^{n \Psi_{+,0}} = 0,
\ee
which has a solution:
\be
\Psi_{0+} (z) = - \frac{2}{n} \ln z + \frac{1}{n} \ln \frac{2(m+n)}{n}.
\label{Psi_0+}
\ee
For a negatively charged plate, Eq.~(\ref{PB-Psi}) reduces to
 \be
- \Psi_{0-}''(z) - \frac{1}{m+n} e^{-m \Psi_{-,0}} = 0,
\ee
which has a solution:
\be
\Psi_{0-} (z) =  \frac{2}{m} \ln z - \frac{1}{m} \ln \frac{2(m+n)}{m}.
\label{Psi_0-}
\ee

The most salient feature of these solutions are their singularities at $z = 0$.  The potential diverges logarithmically, while ion density diverges as $z^{-2}$ as $z \rightarrow 0$.  Therefore Eqs.~(\ref{Psi_0+},\ref{Psi_0-}) are the potentials generated by an {\em infinitely} charged plate located at the origin in the presence of counter-ions only.  However, since the co-ions are strongly repelled by the plate anyway, their existence is not going to change the nature of the singularity at $z = 0$.  The fact that the potential generated by an infinitely positive/negative charged surface inside an electrolyte is finite for any $z>0$ is probably the most important property of Poisson-Boltzmann theory.  It was named (somewhat vaguely) as {\em charge renormalization} by Alexander {\it et. al.} \cite{charge-RG-Alexander-JCP-1984}.   This property of course can only be approximately applicable in real systems.  All ions have finite volumes and the density of screening ions must saturate at the corresponding close packing value.  For example, inside a $1mM$ monovalent salt and near a surface with potential $0.3V$, the ionic concentration would be about $10^5mM$ according to the Poisson Boltzmann theory, which already exceeds the close packing limit, if every ion has (hydrated) volume of $30\AA^3$.  The Poisson-Boltzmann theory can be modified to incorporate finite size effects of ions \cite{Borukhov:1997fk,Zhou:2011uq}.  


\begin{center}
\begin{table*}[!hbt]
\begin{tabular}{ccl}
\hline\hline \noalign{\smallskip}
Electrolytes \,\,
& Plate \,\,& Near field expansion of $\theta_{\pm}$
\\  \noalign{\smallskip}\hline \noalign{\smallskip}
\multirow{2}{*} {$3:-1$} & $+$ &
$\theta_+ = \sum_{k =1} a_{2k}     z^{2k}$ \vspace{1mm}\\
& $-$ & $\theta_- =  \sum_{k=0} a_{2+2k/3} z^{2+2k/3}$
\\ \noalign{\smallskip} \hline \noalign{\smallskip}
\multirow{2}{*} {$4:-1$} & $+$ &
$\theta_+ = \sum_{k =1} a_{2k} z^{2k}$ \vspace{3mm}\\
& $-$ & $\theta_- =  \sum_{k=0} a_{2+k/2} z^{2+k/2}$
\\\noalign{\smallskip} \hline \noalign{\smallskip}
\multirow{2}{*} {$3:-2$} & $+$ &
$\theta_+ = \sum_{k =2} a_{k} z^{k}+z^{3}\sum_{k=1}b_{2k}z^{2k}$ \vspace{3mm}\\
& $-$ & $\theta_- =  \sum_{k=1} a_{2k} z^{2k }
+ z^{4/3} \sum_{k=1} b_{2k} z^{2k} + z^{8/3} \sum_{k=1}c_{2k} z^{2k} $
\\\noalign{\smallskip} \hline \noalign{\smallskip}
\multirow{2}{*} {$4:-3$} & $+$ &
$\theta_+ = \sum_{k=1} a_{2k} z^{2k }
+ z^{8/3} \sum_{k=1} b_{2k} z^{2k}
+ z^{16/3} \sum_{k=1} c_{2k} z^{2k} $ \vspace{3mm}\\
& $-$ & $\theta_- =  \sum_{k=1} a_{2k} z^{2k }
+ z^{3/2} \sum_{k=1} b_{2k} z^{2k} + z^3 \sum_{k=1}c_{2k} z^{2k}+ z^{9/2} \sum_{k=1} d_{2k} z^{2k} $
\\\noalign{\smallskip} \hline \noalign{\smallskip}
\end{tabular}
\caption{ Near field expansion of $\theta_{\pm}$ as defined in Eq.~(\ref{eta-def}), where $\Psi_{0\pm}(z) $ are defined in Eqs.~(\ref{Psi_0+},\ref{Psi_0-}).  Note the terms with fractional powers of $z$ in the expansions.   All coefficients are determined by the lowest order one $a_2$, which is shown in Eq.~(\ref{a_2}).  }
\label{Table:eta-expansion}
\end{table*}
\end{center}

From now on we shall fix the infinitely charged plate at the origin.  Eqs.~(\ref{Psi_0+},\ref{Psi_0-}) are the leading terms of systematic near field expansions for the solutions to the full PB equation.   Defining
\be
\Psi_{\pm}(z) = \Psi_{0\pm}(z) + \theta_{\pm}(z),
\label{eta-def}
\ee
and substitute these back into the full PB equation  Eq.~(\ref{first-integral}), we obtain the equations for $\theta_{\pm}(z)$:
\begin{widetext}
\begin{subequations}
\label{eta-pm}
\bea
- \frac 1{2}\, {\theta_+'} ^2 + \frac {2}{n z}\,\theta_+ + \frac{1}{m+n} \left[
\frac{2(m+n)}{n^2 z^2}\, (e^{n\theta_+} - 1)
-\frac 1{m} \left( \frac{n}{2(m+n)} \right)^{\frac{m}{n}}
z^{\frac{2m}{n}} e^{-m \theta_+} -\frac 1{n}-\frac 1{m} \right] &=& 0,
\\ \noalign{\smallskip}\noalign{\smallskip}
- \frac 1{2}\, {\theta_-'} ^2 - \frac {2}{m z}\,\theta_- + \frac{1}{m+n} \left[\frac 1{n} \left( \frac{m}{2(m+n)} \right)^{\frac{n}{m}}
z^{\frac{2n}{m}} e^{n \theta_-}
+\frac{2(m+n)}{m^2 z^2}\, (e^{-m\theta_-}-1)
-\frac 1{n}-\frac 1{m} \right] &=& 0.
\eea
\end{subequations}
\end{widetext}%
Using method of dominant balance \cite{Bender-Orszag}, it is easy to see that for small $z$ the leading terms for $\theta_{\pm}$ scale as $z^2$.  However, it would be incorrect to conclude that $\theta_{\pm}$ can be expanded into Taylor series of $z$.   For non-integer $2m/n$ ($2n/m$), the equation for $\theta_+$ ($\theta_-$)  contains nonanalytic term $z^{2m/n}$ ($z^{2n/m}$), and hence the corresponding expansion of $\theta_+$ ($\theta_-$) must also contain this non-analytic term, as well as its integer powers.   The explicit forms of the near field expansion for various cases of electrolytes are displayed in Table \ref{Table:eta-expansion}.  Using Wolfram Mathematica and  substituting these expansions into the equations for $\theta_{\pm}$, Eqs.~(\ref{eta-pm}), all coefficients are determined in terms of $a_2$, which in turn can be determined exactly:
\bea
-a^+_2 =  \frac{1}{6m},\quad
a^-_2 =  \frac{1}{6n}.
\label{a_2}
\eea
The near field expansion therefore does not contain any undetermined parameter.   We summarize the coefficients in the Appendix \ref{Table:near-field-coefficient}.



\section{Far Field Asymptotics}
\label{sec:far}
The simple form Eq.~(\ref{far-field-general}) is only the leading order term in the far field expansion.  To find all higher order terms,  it is convenient to introduce the following variables:
\be
s = e^{-z}, \quad \Upsilon(s) = e^{\Psi(z)}.
\label{Psi-Upsilon}
\ee
Substituting these into the PB Eq.~(\ref{first-integral}) we find:
\be
-\frac 1{2} \left( \frac {s\Upsilon'} {\Upsilon}\right)^2
+\frac {1}{m+n}\left( \frac {\Upsilon^n-1}{n}+\frac{\Upsilon^{-m}-1}{m} \right)= 0.
\label{PB-Upsilon}
\ee
The far field limit $z \rightarrow \infty$ corresponds to $s = 0$.  Around this point, $\Upsilon(s)$ can be expanded into a Taylor series:
\be
\Upsilon(s) = \sum_{k = 0}^{\infty} c_k \, s^k
= 1+ c_1 \, s + c_2 \, s^2 + \ldots.
\label{far-field-expansion}
\ee
The zeroth order coefficient $c_0 = 1$ because as $z \rightarrow \infty$, $s \rightarrow 0$, $\Psi \rightarrow 0$, and  $\Upsilon = e^{\Psi} \rightarrow 1$.  Substituting the expansion Eq.~(\ref{far-field-expansion}) into Eq.~(\ref{PB-Upsilon}) and compare the coefficients of $s^k$ order by order, we easily obtain relations between $c_k$'s, which allow us to solve all $c_k$'s in terms of $c_1$:
\be
c_k = \hat{c}_k \, c_1^k.
\label{coeffients-far-field}
\ee
The numbers $\hat{c}_k$ can be calculated straightforwardly using Wolfram Mathematica up to arbitrary order.  In this work we only keep twelve terms  for $(m: -1)(m = 1,2,3,4)$ electrolytes, and sixteen terms for $(3: -2)$ and $(4: -3)$ cases.
Some of these coefficients $\hat{c}_k$ are tabulated in Table \ref{Table:c_hat}.  The only remaining coefficient $c_1$ will be determined by matching the far field expansion with the near field expansion, which, by construction, is due to an infinitely charged plate at the origin.

Combining Eq.~(\ref{far-field-expansion}) and Eq.~(\ref{Psi-Upsilon}) we can also obtain the far field expansion for the potential $\Psi$.  To the leading order we have:
\be
\Psi(z) = c_1 \,e^{-z} + O(e^{-2z}).
\ee
Comparing this with Eq.~(\ref{far-field-general}) and Eq.~(\ref{eta-sigma-R}) (also noticing the relation between $\phi$ and $\Psi$) we see that
\be
c_1 = \eta_R(\pm \infty),
\ee
i.e., $c_1$ is precisely the dimensionless renormalized dimensionless surface charge density for an infinitely charged plate.  

\begin{center}
\begin{table*}[!hbt]
\begin{tabular}{ccccccccccccccccc}
\hline\hline \noalign{\smallskip}
 	& \,\, $\hat{c}_2$ \,\, & \,\,  $\hat{c}_3$   \,\, & \,\,  $\hat{c}_4$   \,\,
	& \,\,  $\hat{c}_5$ \,\,   & \,\,  $\hat{c}_6$   \,\, & \,\,  $\hat{c}_7$   \,\,
	& \,\,  $\hat{c}_8$   \,\, & $\hat{c}_9$ & $\hat{c}_{10}$
	& $\hat{c}_{11}$  & $\hat{c}_{12}$
\\ \noalign{\smallskip} \hline\noalign{\smallskip}
$1:-1$ & $\frac{1}{2}$ & $\frac{3}{16}$  & $\frac{1}{16}$ & $\frac{5}{256}$
& $\frac{3}{512}$ & $\frac{7}{4096}$  & $\frac{1}{2048}$ & $\frac{9}{65536}$
 & $\frac{5}{131072}$ & $\frac{11}{1048576}$  & $\frac{3}{1048576}$
 \\ \noalign{\smallskip}\noalign{\smallskip}\hline\noalign{\smallskip}
$2:-1$ & $\frac{1}{3}$ & $\frac{1}{12}$  & $\frac{1}{54}$ & $\frac{5}{1296}$
& $\frac{1}{1296}$ & $\frac{7}{46656}$  & $\frac{1}{34992}$ & $\frac{1}{186624}$
 & $\frac{5}{5038848}$ & $\frac{11}{60466176}$  & $\frac{1}{30233088}$
 \\ \noalign{\smallskip} \noalign{\smallskip} \hline\noalign{\smallskip}
$3:-1$ & $\frac{1}{6}$ & $\frac{1}{16}$  & $ -\frac{5}{432}$ & $\frac{269}{20736}$
& $-\frac{125}{13824}$ & $\frac{21469}{2985984}$  &$-\frac{25997}{4478976}$
 & $\frac{76699}{15925248}$ & $-\frac{10511275}{61917364224}$  &$\frac{216900209}{61917364224}$ &$-\frac{693028429}{227030335488}$
  \\ \noalign{\smallskip}\noalign{\smallskip}\hline\noalign{\smallskip}
  $4:-1$ & $0$ & $\frac{1}{8}$  & $ -\frac{1}{10}$ & $\frac{33}{320}$
& $-\frac{11}{100}$ & $\frac{313}{2560}$  &$-\frac{1567}{11200}$
 & $\frac{83853}{512000}$ & $-\frac{1749}{8960}$  &$\frac{33847223}{143360000}$ &$-\frac{22792479}{78848000}$
  \\ \noalign{\smallskip}\noalign{\smallskip}\hline\noalign{\smallskip}
 $3:-2$ & $\frac{1}{3}$ & $\frac{1}{6}$  & $ \frac{7}{135}$ & $\frac{49}{1620}$
& $\frac{53}{8100}$ & $\frac{941}{145800}$  &$\frac{1}{54675}$
 & $\frac{673}{364500}$ & $-\frac{4549}{7873200}$  &$\frac{109487}{147622500}$ &$-\frac{252281}{590490000}$
  \\ \noalign{\smallskip}\noalign{\smallskip}\hline\noalign{\smallskip}
  $4:-3$ & $\frac{1}{3}$ & $\frac{7}{24}$  & $ \frac{11}{108}$ & $\frac{605}{5184}$
& $\frac{935}{36288}$ & $\frac{146399}{2612736}$  &-$\frac{17149}{13716864}$
 & $\frac{4797881}{146313216}$ & $-\frac{6268525 }{564350976}$  &$\frac{2229207493}{94810963968}$ &$-\frac{703428293}{47405481984}$
  \\ \noalign{\smallskip}\noalign{\smallskip}\hline\noalign{\smallskip}
\end{tabular}
\caption{Some coefficients $\hat{c}_k$ of far field expansions. }
\label{Table:c_hat}
\end{table*}
\end{center}


To employ asymptotic matching of the far field expansions and the near field one, we will have to evaluate the truncated far field expansion at  $z^* \sim 1$.   However, It turns out that, for many cases of $(m,-n)$, the far field expansion does not converge near $z^* \sim 1$.  \footnote{Even though the potential $\Psi(z)$, and hence also $\Upsilon(s) = e^{\Psi(z)}$ as well,  is finite for all $s \in (0,1)$, their Taylor expansions around $s = 0$ may diverge. }   We use the well known Shanks transformation \cite{Bender-Orszag} to improve convergence for the far field expansion of $\Upsilon(s)$.  In Appendix~\ref{app:Shanks} we give a brief introduction to this method.  The useful Shanks transformation is of second order.   Consider a power series $A = \sum_n a_n x^n$.  The partial sums is $A_n = \sum_{k \leq n} a_n x^n$, its second order Shank transform is given by
\be
S_2 (A_n) = \frac{
 \left| \begin{array}{ccc}
A_{n-2}& A_{n-1} & A_n \\
\Delta A_{n-2} & \Delta A_{n-1} & \Delta A_n\\
\Delta A_{n-1} & \Delta A_n & \Delta A_{n+1}
\end{array}\right| }{
\left| \begin{array}{ccc}
1 &1 & 1 \\
\Delta A_{n-2} & \Delta A_{n-1} & \Delta A_n\\
\Delta A_{n-1} & \Delta A_n & \Delta A_{n+1}
\end{array}\right| },
\ee
where $\Delta A_n = A_n - A_{n-1} = a_n x^n$.  Replacing of $A_n$ by $S_2(A_n)$ may lead to substantial improvement of convergence. 

Implementation of the second order Shanks transform is straightforward using Wolfram Mathematica.  For the cases of $1:-1$ and $2:-1$, Shanks transform yields very simple result:
\be
\Upsilon_{\rm Shank}(s) = \left\{
\begin{array}{ll}
\frac{(4 + c_1 \, s)^2}
{(4 - c_1 \, s)^2},
& \quad 1:-1;
\vspace{4mm}\\
\frac{36 + 24 c_1 s + c_1^2 s^2}
{(6-c_1 s)^2},
& \quad 2:-1.
\end{array}
\right.
\ee
which exactly solve Eq.~(\ref{PB-Upsilon}).  They correspond to {\em exact solutions} to the original nonlinear PB equation Eq.~(\ref{first-integral}):
\be
\Psi(z) = \left\{
\begin{array}{ll}
2 \ln \frac{(4 + c_1 e^{-z})}
{(4 - c_1 e^{-z})},
& \quad 1:-1;
\vspace{4mm}\\
\ln \frac{36 + 24 \, c_1  e^{-z} + c_1^2 e^{-2z}}
{(6-c_1 e^{-z})^2},
& \quad 2:-1.
\end{array}
\right.
\ee
The parameter $c_1$ can be determined by matching with the boundary conditions on the plate.   It is interesting to note that in the far field expansions $c_1$ always appear together with $e^{-z}$, hence change of $c_1$ amounts to translation of the coordinate $z$.

\begin{figure}
\begin{center}
\includegraphics[width=6.5cm]{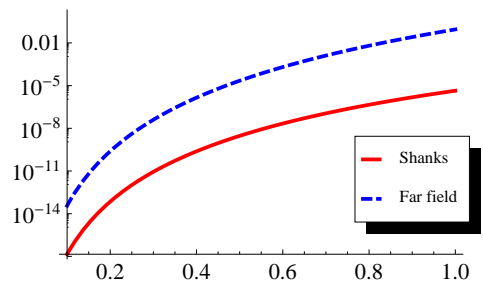}
\caption{The error of truncated far field expansion as an approximate solution to the nonlinear PB equation Eq.~(\ref{PB-Upsilon}), as well as that of the Shanks transformed version, for the case of $(3:-1)$.   The Shanks transform reduces the error by about three orders of magnitude.   }
\label{error-PB-shanks}
\end{center}
\vspace{-5mm}
\end{figure}

For other valences $m:-n$, the Shanks transforms of the far field expansions do not yield exact solutions.  Nevertheless, they remarkably improve the quality of approximation.  We  substitute the far field expansions (with $c_1$ set to unity) as well as their Shanks transformed versions into the nonlinear PB equation Eq.~(\ref{PB-Upsilon}).  Since they are not exact solutions, the right hand sides do not identically vanish, but measure the quality of the truncated far field expansions.  As shown in Fig.~\ref{error-PB-shanks}, the Shanks transform reduces the error of the far field expansion by about three orders of magnitude for the case of $(3:-1)$.  Similar improvements are also achieved for other cases.

\section{Asymptotic Matching}
\label{sec:matching}

\begin{figure}[htb]
\begin{center}
\includegraphics[width=6cm]{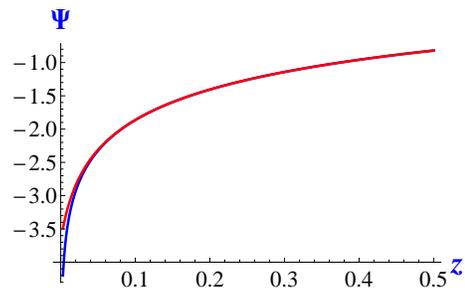}
\caption{Matching of near field and far field expansions for a negatively charged plate in $3:-1$ electrolyte. Top: far-field expansion, Bottom: near field expansion. Evidently, the far field expansion is excellent even around tenth of Debye length.  }
\label{Fig:matching}
\end{center}
\vspace{-5mm}
\end{figure}

\begin{center}
\begin{table*}[!hbt]
\begin{tabular}{ccrccrrr}
\hline\hline \noalign{\medskip}
electrolyte & \,plate\, \,\,&  \,\, $z^*$ &\quad $N_{\rm far}\quad$&\quad$N_{\rm near}$&\,$a_2 $ &\quad $c_1 = \eta_R(\pm \infty)$\,\, & \quad\quad\quad  $\delta c_1$ \\
\noalign{\smallskip} \hline \noalign{\smallskip}
\multirow{2}{*} {$1:-1$} & $+$ &/&12& / & / & $4$& exact \\
& $-$ &/&12& / &/& $-4$& exact
\\\noalign{\smallskip} \hline \noalign{\smallskip}
\multirow{2}{*} {$m:-m$} & $+$ &/&12& $/$ & / & $4/m$ & exact \\
& $-$ & / &12&/&/& $-4/m$& exact
\\\noalign{\smallskip} \hline \noalign{\smallskip}
\multirow{2}{*} {$2:-1$} & $+$ & /&12&/& / &  $6$& exact \\
& $-$ & / &12&/&/& $-6(2 - \sqrt{3})$& exact
\\\noalign{\smallskip} \hline \noalign{\smallskip}
\multirow{2}{*} {$1:-2$} & $+$ & /&12&/ & / & $6(2 - \sqrt{3})$& exact \\
& $-$ & / &12&/&/&$-6$& exact
\\\noalign{\smallskip} \hline \noalign{\smallskip}
\multirow{2}{*} {$2m:-m$} & $+$ &/&12&/& / & $6/m$ & exact \\
& $-$ & / &12&/&/& $-6(2 - \sqrt{3})/m$& exact
\\\noalign{\smallskip} \hline \noalign{\smallskip}
\multirow{2}{*} {$3:-1$} & $+$
& 2.0&12&11& 1/18 & 8.707001 & $8\times10^{-6}$ \\
& $-$ & $0.5$&12&10 & -1/6 & $-0.9938796$ & $2\times10^{-7}$
\\\noalign{\smallskip} \hline \noalign{\smallskip}
\multirow{2}{*} {$1:-3$} & $+$
& 0.5&12& 10& 1/6 &0.9938796& $2\times10^{-7}$ \\
& $-$ & $2.0$ & 12&11& -1/18 & $-8.707001$ &$8\times10^{-6}$
\\\noalign{\smallskip} \hline \noalign{\smallskip}
\multirow{2}{*} {$4:-1$} & $+$
& 2.5 &12&11& 1/24 & 12.3142 & $0.0002$ \\
& $-$ &0.4& 12&10 &-1/6 &-0.717417& $2\times10^{-6}$
\\\noalign{\smallskip} \hline \noalign{\smallskip}
\multirow{2}{*} {$3:-2$} & $+$
& 0.7 &16&13& 1/18 & 2.45953455 & $1\times 10^{-8}$ \\
& $-$ & 0.6 &16&9& -1/12 & -1.1542225 & $3\times 10^{-7}$
\\\noalign{\smallskip} \hline \noalign{\smallskip}
\multirow{2}{*} {$4:-3$} & $+$
& 1.0 &16&13& 1/24 & 1.53280486 & $1\times 10^{-8}$ \\
& $-$ &0.6 & 16&7& -1/18 & -0.89788695 & $7\times 10^{-8}$
\\\noalign{\smallskip} \hline \noalign{\smallskip}
\end{tabular}
\caption{Saturated values of renormalized surface charge density $\eta_{R}(\pm \infty)$ for different electrolytes.  $N_{\rm far}$ and $N_{\rm near}$ are numbers of terms kept in the far field and near field expansions.  \quad \quad Two identities that can be used to generalize the above results to other cases:
$\eta_R^{m:-n}(\infty) = - \eta_R^{n:-m}(-\infty)$, and
$\eta_R^{m:-n}(\infty) = p^{-1} \eta_R^{\tilde{m}:-\tilde{n}}(\infty)$, where
$m = p \tilde{m}, n = p \tilde{n}$. }
\label{Table:c1}
\end{table*}
\end{center}

We now require that the far field expansions equal to the near field expansions (both of them are, of course, approximate) at some intermediate point $z^*$:
 \bea
  \Psi^{\rm app}_{\rm far}(z^*,c_1)
 =  \Psi^{\rm app}_{\rm near}(z^*).
 \label{matching-condition}
 \eea
Note that only the far field expansion depends on $c_1$.   Solving this equation determines the only parameter in the far field expansion, i.e. the renormalized charge density $c_1 = \eta_{R}(\pm \infty)$.  The matching condition Eq.~(\ref{matching-condition}) is numerically solved for $c_1$ using Mathematica.   The solution however can only be an approximation to the exact value $c_1^{\rm ex}$, since both near field and far field expansions are approximate. The errors of these expansions can be estimated by the difference between the current order approximation and the next order approximation.
Via some simple analysis, to be detailed in Appendix \ref{app:error}, we derive the following approximate bound for the error  $\delta c_1 = c_1 - c_1^{\rm ex}$:
\be
|\delta c_1| \leq
	\frac{|\delta \Psi^{\rm app}_{\rm near}(z^*)|
+ |\delta \Psi^{\rm app}_{\rm far}(z^*,c_1)|}
{ |\frac{\partial \Psi^{\rm app}_{\rm far}(z^*,c_1)}{\partial c_1}|}
\ee
We systematically vary the matching point $z^*$ to minimize this error bound.  The optimal $z^*$, the calculated coefficient $c_1$, and the corresponding error bound for all cases are shown on the Table.~\ref{Table:c1}.  Except for the case of positively charged plate in $4:-1$ electrolyte, the errors are all with in orders of $10^{-6}$.
Also shown there are the number of terms used in the near field and far field expansions.   In Fig.~\ref{Fig:matching} we show both near field and far field expansions for the case of a negatively charged plate  in $3:-1$ electrolyte.  It can be seen there that the Shank transformed far field expansion is excellent even if $z^*$ is within a Debye length to the plate.

Note that our asymptotic matching is different from the one discussed in standard textbooks of asymptotic analysis.  There, it is required that there is a controlling small parameter $\epsilon$, and a whole intermediate region where both far field and near field expansions become exact as $\epsilon \rightarrow 0$.   It is then guaranteed that in this limit, the matching becomes exact.  In our problem, there is no small parameter whatsoever.  The precision of both far field and near field expansions are guaranteed by the high order of expansions, augmented by Shanks transformation.  Evidently, analysis with these high order expansions would be extremely time consuming without modern computer softwares for symbolic calculation, such as Wolfram Mathematica or Maple.



\section{Renormalized Charge Density of a Highly Charged Plate}
\label{sec:highly-charged}

\begin{center}
\begin{table*}[!thb]
\begin{tabular}{ccl}
\hline\hline \noalign{\smallskip}
Electrolytes \,\,
& Plate \,\,&Asymptotic expansion of $\eta_R(\eta)$
\\  \noalign{\smallskip}\hline \noalign{\smallskip}
$1:-1$
& $\pm$ &
$ \frac{2 \eta}{1+ \sqrt{1+(\eta/2)^2}}$ (exact)
\\ \noalign{\smallskip} \hline \noalign{\smallskip}
\multirow{2}{*}
{$2:-1$}\footnote{The exact result is given implicitly by
$\eta = \frac{36 \eta _R \left(\eta_R+6\right)}
 {\left(6-\eta _R\right)  \left(\eta _R^2 + 24 \eta_R +36\right)}.  $}
 & $+$ &
$ 6
   -\frac{12}{\eta}
   +\frac{12}{\eta^2}
   -\frac{4}{\eta^3}
   -\frac{4}{\eta^4}
   +\frac{4}{\eta^5}
   +\frac{4}{3 \eta^6}
   -\frac{4}{\eta^7}
   +\frac{4}{3 \eta^8}
   +\frac{92}{27 \eta^9}
   -\frac{124}{27 \eta^{10}}
   +O\left( \eta^{-11} \right)$
   \vspace{1mm}\\
& $-$ &
$  6\left(\sqrt{3}-2\right)
   +\frac{6\left(\sqrt{3}-2\right)}{\eta}
   +\frac{3\left(\sqrt{3}-2\right)}{\eta^2}
   +\frac{2-\sqrt{3}}{\eta^3}
   +\frac{2-\frac{3\sqrt{3}}{4}}{\eta^4}
   +\frac{\frac{5\sqrt{3}}{4}-2}{\eta^5}
   +\frac{\sqrt{3}-16}{24
   \eta^6}
   +O\left(\eta^{-7} \right)$
 \\ \noalign{\smallskip} \hline \noalign{\smallskip}
\multirow{2}{*}
{$3:-1$}
& $+$ &
$ 8.70701\left( 1
 - \frac {2}{\eta}
 +\frac {2}{\eta^2}
 -\frac{0.888888}{\eta^3}
 -\frac{0.222225}{\eta^4}
+\frac{0.444446}{\eta^5}
-\frac{0.0493818}{\eta^6}
-\frac{0.197532}{\eta^7}
\right)
+ O\left(\eta^{-8} \right)$
 \vspace{1mm}\\
& $-$ &
$ -0.993871\left(1
-\frac{2}{3|\eta|}
+\frac{2}{9|\eta|^2}
+\frac{0.0987202}{|\eta|^3}
-\frac{0.0667656}{|\eta|^{11/3}}
-\frac{0.0905049}{|\eta|^4}
+\frac{0.0444704}{|\eta|^{14/3}}
-\frac{0.0274087}{|\eta|^5}\right)
+O\left(|\eta|^{-17/3}\right)$
 \\ \noalign{\smallskip} \hline \noalign{\smallskip}
\multirow{2}{*}
{$4:-1$}
& $+$ &
$12.3141\left(1
 - \frac {2}{\eta}
 +\frac {2}{\eta^2}
 -\frac{0.999991}{\eta^3}
 -\frac{0.0000187721}{\eta^4}
 +\frac{0.300013}{\eta^5}
 -\frac{0.0999981}{\eta^6}
 -\frac{0.0785797}{\eta^7}\right)
 +O\left(\eta^{-8}\right)$
 \vspace{1mm}\\
& $-$ &
$0.717419\left(1
 - \frac {2}{|\eta|}
 +\frac {1}{8|\eta|^2}
 +\frac{0.0625227}{|\eta|^3}
 -\frac{0.0401672}{|\eta|^{7/2}}
 -\frac{0.0390738}{|\eta|^4}
 +\frac{0.0200836}{|\eta|^{9/2}}
 -\frac{0.0148545}{|\eta|^5}\right)
 +O\left(\eta^{-11/2}\right)$
 \\ \noalign{\smallskip} \hline \noalign{\smallskip}
\multirow{2}{*}
{$3:-2$}
& $+$ &
$2.46097\left(1
 - \frac {2}{\eta}
 +\frac {2}{\eta^2}
 -\frac {0.0554251}{\eta^3}
 -\frac{0.0695749}{\eta^4}
 +\frac{0.0250130}{\eta^5}
 -\frac{0.00634138}{\eta^6}
 -\frac{0.00260921}{\eta^7}\right)
 +O\left(\eta^{-8}\right)$
  \vspace{1mm}\\
& $-$ &
$-1.154222\left(1
-\frac{2}{3|\eta|}
+\frac{2}{9|\eta|^2}
+\frac{0.0252192}{|\eta|^3}
-\frac{0.0415042}{|\eta|^4}
-\frac{0.0133842}{|\eta|^{13/3}}
+\frac{0.000454105}{|\eta|^5}
+\frac{0.00892278}{|\eta|^{16/3}}\right)
+O\left(|\eta|^{-6}\right)$
 \\ \noalign{\smallskip} \hline \noalign{\smallskip}
\multirow{2}{*}
{$4:-3$}
& $+$ &
$1.53278\left(1
 - \frac {2}{3\eta}
 +\frac {2}{9\eta^2}
 -\frac {0.0123480}{\eta^3}
 -\frac{0.0164594}{\eta^4}
 +\frac{0.00342930}{\eta^5}
 -\frac{0.00085279}{\eta^{17/3}}
 +\frac{0.00144768}{\eta^6}\right)
 +O\left(\eta^{-20/3} \right)$
 \vspace{1mm}\\
 & $-$ &
 $-0.897888\left(1
 -\frac{1}{2|\eta|}
 +\frac{1}{8|\eta|^2}
 +\frac{0.00694196}{|\eta|^3}
 -\frac{0.0112835}{|\eta|^4}
 -\frac{0.00255865}{|\eta|^{9/2}}
 +\frac{0.000434214}{|\eta|^5}
 +\frac{0.00127932}{|\eta|^{11/2}}\right)
 +O\left(|\eta|^{-6}\right)$
\\ \hline\hline \noalign{\smallskip}
\end{tabular}
\caption{Asymptotic expansions of the renormalized dimensionless surface charge densities $\eta_R(\eta)$ in terms of the bare dimensionless density $\eta$, for different electrolytes. Using  Eq.~(\ref{eta_pm}, \ref{p}), we can generalize these to other cases as well. }
\label{Table:eta-R-expansion}
\end{table*}
\end{center}

\begin{figure*}[htb!]
\begin{center}
\includegraphics[width=6cm]{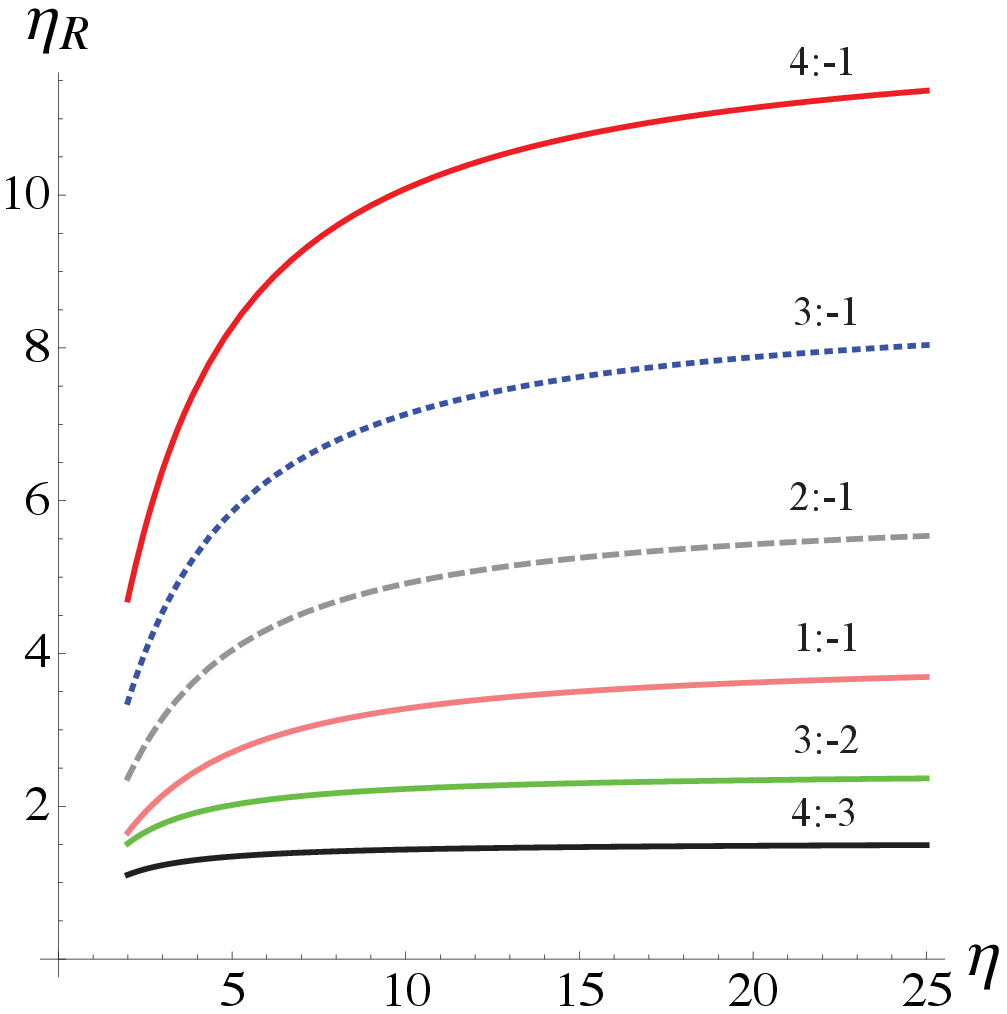}
\hspace{5mm}
\includegraphics[width=6cm]{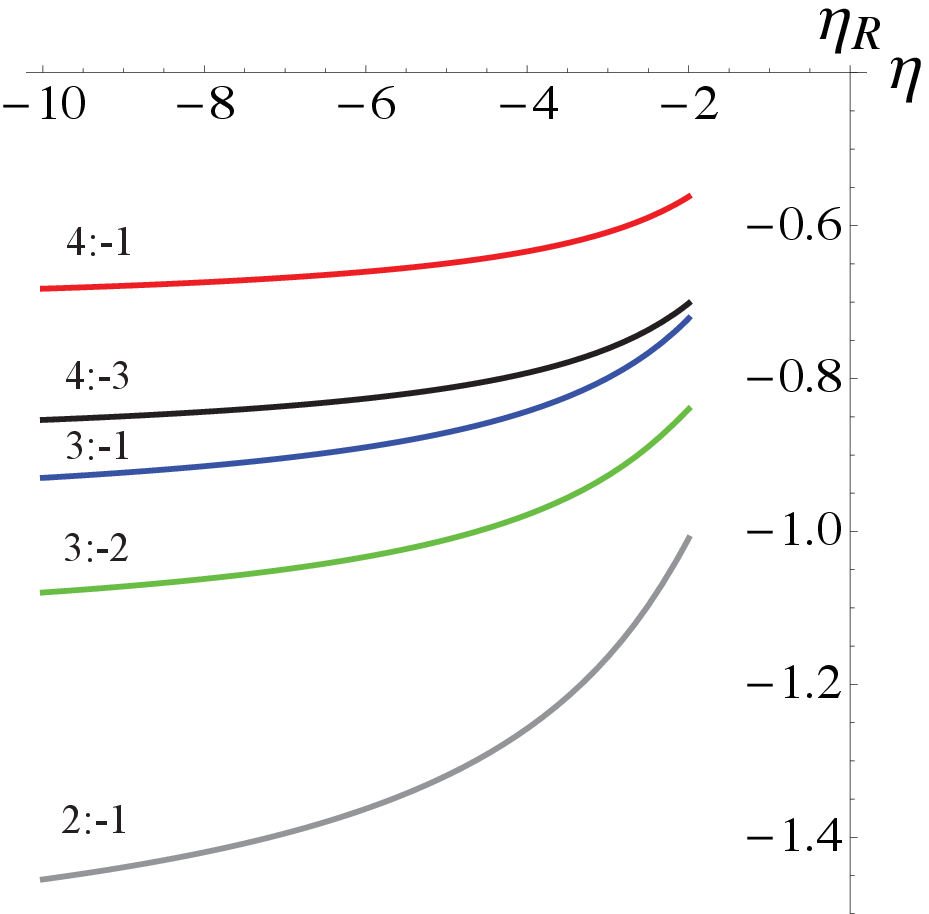}
\caption{Renormalized surface charge densities as functions of bare surface charge densities.
Top: positively charged plate. Bottom: negatively charged plate.  }
\label{Fig:renormalized-charge}
\end{center}
\vspace{-5mm}
\end{figure*}

Using the calculated coefficient $c_1 = \eta_R(\pm \infty)$, we can now completely determine the far field asymptotic expansion of the potential $\Psi(z)$.   Recall the near field expansion of $\Psi(z)$ exhibits a logarithmic singularity at $z = 0$, see Eqs.~(\ref{Psi_0+},\ref{Psi_0-}), which corresponds to an infinitely charged plate located at the origin.  What we really want is however the asymptotics of a finitely charged plate, say, with dimensionless surface charge density $\eta$.   We can always set up the coordinate system such that the $\eta$ plate is located at $z_0$ (cf Eq.~(\ref{Neumann-eta})), such that if we extrapolate the solution $\Psi(z)$ to the region $z < z_0$, there is a singularity at $z=0$.  This means that the potential $\Psi(z)$ is identical to the one have analyzed in previously.  The location $z_0$ of the plate now becomes a function of $\eta$, through the Neumann boundary condition Eq.~(\ref{Neumann-eta}).   Using the near field expansions for $\Psi$, the boundary condition Eq.~(\ref{Neumann-eta}) now becomes
\be
- \frac{d}{dz} \Psi_{\rm near}(z_0) = \eta.
\ee
The left hand side can be expanded into an asymptotic series in $z_0$, using the near field expansion discussed in Sec.~\ref{sec:near}.  The series can be inverted to yield $z_0$ as an asymptotic expansion in terms of $\eta = \ell_{DB}/\mu$, which is also small for strongly charged surface.  The results are however to complicated to be shown explicitly.

On the other hand, the leading order far field asymptotics can be written as
\be
\Psi_{\rm far}(z) \sim \eta_R(\pm \infty)\, e^{-z}  =  \eta_R(\pm \infty) \,e^{-z_0} \, e^{-\delta z},
\ee
where $\delta z = z - z_0$ is {\em the distance from the plate to the field point}.  Therefore by the definition Eq.~(\ref{far-field-general}), the renormalized dimensionless surface charge density of the plate with bare surface charge $\eta$ is given by
\be
\eta_R(\eta) =  \eta_R(\pm \infty) \, e^{-z_0}.
\ee
Now expanding $e^{-z_0}$ into asymptotic series of $\eta$, we finally obtain the renormalized surface charge density $\eta_R(\eta)$ as an asymptotic series in terms of the bare surface charge density $\eta$. The results are shown  in Table \ref{Table:eta-R-expansion} and Fig.~\ref{Fig:renormalized-charge}.   Even though these results are derived for strongly charged surface, they are quantitatively accurate even for weakly charged surfaces.  For example, for the case of  $\eta = \ell_{DB}/\mu = 5$, the relative error for $1:-1$ electrolyte is of order of $10^{-6}$.

\section{Concluding Remarks}
In this work, we have developed systematic near field and far field expansions for  the electrostatic potential generated by a strongly charged plate inside a generic charge-asymmetric electrolyte. Using these expansions,  we have derived a perturbative expansion for the renormalized surface charge density  in terms of the bare surface charge density.   Our methods can be immediately generalized to charged plates immersed in mixed electrolytes. 

The authors thanks Shanghai Jiao Tong University and NSFC (grant numbers 11174196, 91130012) for financial support.



\appendix
\section{Basics of Shanks Transform}
\label{app:Shanks}
In this appendix, we present a heuristic discussion of Shanks transform.  The readers are referred to the classical monograph by Bender and Orszag  \cite{Bender-Orszag} for details.  Originally Shanks transformation was invented to speed up convergence of slowly converging series.   Its most powerful application is however to sum up divergent series and yield finite result.  For simple geometric series, Shanks transformation is equivalent to the traditional procedure, where one first sum up the series within its radius of convergence, and then analytically continuate beyond this domain.

Consider a simple geometric series
\be
A = \sum_{0}^{\infty} a\, \left( \frac{z}{z_1}\right)^k,
\label{geometric-series}
\ee
which is convergent for $|z|< z_1$.  Within this domain of convergence, the series can be summed up to yield a simple rational function with one simple pole at $z = z_1$:
\be
A = \frac{a}{1-z/z_1}.
\label{A-simple-pole}
\ee
Let us define the remainder $R_n$ as
\be
R_n = A - A_n.
\ee
which is the error if we use the partial sum $A_n$ to approximate the original series.
For the simple geometric series Eq.~(\ref{geometric-series}) the remainder has the following simple form
\be
R_n = \frac{a}{1-z/z_1} \left( \frac{z}{z_1}\right)^n,
\label{remainder-geometric}
\ee
which is also geometric in $n$.  The remainder goes to zero as long as the original series converges.

Suppose we are interested in the domain $|z|>1$, i.e. outside the radius of convergence of $A$, and suppose that for some reason, we are only able to calculate the partial sums $A_n = \sum_0^n a\, z^k$ only up to some finite order $n$.  The conventional approach fails here because we can not sum up the series to infinite order for small $z$.  On the other hand, the partial sum $A_n$ is not a good approximation to the exact because the error (i.e. the remainder $R_n$) diverges in the domain of interest.  So the question is how we can obtain a good approximation to the series outside its domain of convergence, with the knowledge of only finite terms of the series?

For the geometric series Eq.~(\ref{geometric-series}), we have the following relations for arbitrary integer $n$:
\bea
A_{n+1} &=&  A - \alpha \, z^{n+1}, \\
A_{n} &=&  A - \alpha \, z^{n}, \\
A_{n-1} &=&  A - \alpha \, z^{n-1}.
\eea
Solving these three equations, we can express $A$ in terms of three parameters
$A_{n+1},A_{n},A_{n-1}$ (details ignored here):
\bea
A &=& \left \vert \begin{array}{cc}
A_n & A_{n-1} \\
\Delta A_n & \Delta A_{n-1}
\end{array} \right \vert
{\Bigg /}
\left \vert \begin{array}{cc}
1 & 1 \\
\Delta A_n & \Delta A_{n-1}
\end{array} \right \vert
\nonumber
\\
&= &
\frac{A_{n+1} A_{n-1} - A_n^2}
{A_{n+1} + A_{n-1} - 2 A_n},
\eea
where $\Delta A_n = A_{n+1} - A_n$.  Therefore knowing that the series is geometric, we only need three terms of partial sums to obtain the exact sum of the series.

The Shanks transformation is a nonlinear transformation acting on the partial sums
$S_1(A_n)$ (The subscript $1$ means it is the first order Shanks transformation).
\be
S_1 (A_n) = \frac{A_{n+1} A_{n-1} - A_n^2}
{A_{n+1} + A_{n-1} - 2 A_n}.
\ee
Note however, in computing $S_1(A_n)$, we actually need three partial sums $A_{n+1},A_{n},A_{n-1}$.  Strictly speaking, therefore, Shanks transformation acts on the whole sequence of partial sums, and return a new sequence of partial sums.
It is defined such that it returns the exact $A$ if the original series is geometric.

The utility of Shanks transformation can be best understood by looking at its effect on the remainder.  Let $A_n = A - R_n$ in the above equation, where $R_n$ is the remainder of the original series.  Further let $R^S_n$ be the remainder of the Shanks transformed series:
\be
S_1(A_n) = A - {R}^S_n.
\ee
We easily see the following
\be
R^S_n = \frac{R_{n+1} R_{n-1} - R_n^2}
{R_{n+1} + R_{n-1} - 2 R_n}.
\label{R-S-n}
\ee
If the original series is geometric, then $R_n$ has the form of Eq.~(\ref{remainder-geometric}), and $R^S_n$ simply vanishes, indicating that the transformed partial sum is actually the exact result.

For a more general series, let us assume that we can separate the dominant part of the remainder which is again gemoetric:
\be
R_n= \tilde{R}_n + \alpha \, z^n,
\ee
where $\tilde{R}_n /z^n \rightarrow 0$ as $n \rightarrow \infty$.  Our analysis below applies regardless of the convergence of the remainders $R_n, \tilde{R}_n$.  Using Eq.~(\ref{R-S-n}), the Shanks transformed remainder then becomes:
\bea
R^S_n &=& \frac{a (z \tilde{R}_{n-1} + z^{-1} \tilde{R}_{n+1}
- 2 \tilde{R}_n)}
{a (z + z^{-1} - 2) + (\tilde{R}_{n+1} + \tilde{R}_{n-1} - 2 \tilde{R}_n)/z^n},
\nonumber\\
&\rightarrow& \frac{z \tilde{R}_{n-1} + z^{-1} \tilde{R}_{n+1}
- 2 \tilde{R}_n}{(z + z^{-1} - 2)}.
\eea
In the limit $n \rightarrow \infty$, therefore, the dominant part $z^n$ no longer appear in the transformed version of the remainder.  As a consequence the transformed remainder becomes much smaller than the original remainder.  In particular, if $z>1$ but $\tilde{R}_n \rightarrow 0$, the original series diverges $R_n \rightarrow \infty$ but the Shanks transformed series converges, i.e. $R^S_n \rightarrow 0$.   This shows how Shanks transformation can be used to achieve the purpose of analytic continuation with only finite number of terms of the series.  Note that for the case we studied here, the summation Eq.~(\ref{A-simple-pole}) contains a simple pole at $z=1$.  The effect of Shanks transformation can also be understood as removing this singularity from the original series.   It is easy to see that the Shanks transformation recovers the exacts from Taylor series of all rational functions with only one pole, in the form of $P(z)/(z - z_0)$, where $P(z)$ is a polynomial of arbitrary (but finite) order.

The first order Shanks transformation discussed above does not work if the remainders contain second order poles.
To deal with these series, we have to introduce second order Shanks transformation, which removes two poles at once.  In this case, it can be shown that the exact sum $A$ can be calculated using five terms of the partial sums: $A_{n+2}, A_{n+1},A_n, A_{n-1},A_{n-2}$:
\be
A =  \frac{
\left \vert \begin{array}{ccc}
A_n & A_{n-1} &A_{n-2}\\
\Delta A_n & \Delta A_{n-1}  & \Delta A_{n-2} \\
\Delta A_{n+1} & \Delta A_{n}  & \Delta A_{n-1}
\end{array} \right \vert
} {
\left \vert \begin{array}{ccc}
1 & 1 &1\\
\Delta A_n & \Delta A_{n-1}  & \Delta A_{n-2} \\
\Delta A_{n+1} & \Delta A_{n}  & \Delta A_{n-1}
\end{array} \right \vert
}
\ee

For a general series, the second order Shanks transformation is then defined as
\bea
S_2(A_n) =  \frac{
\left \vert \begin{array}{ccc}
A_n & A_{n-1} &A_{n-2}\\
\Delta A_n & \Delta A_{n-1}  & \Delta A_{n-2} \\
\Delta A_{n+1} & \Delta A_{n}  & \Delta A_{n-1}
\end{array} \right \vert
} {
\left \vert \begin{array}{ccc}
1 & 1 &1\\
\Delta A_n & \Delta A_{n-1}  & \Delta A_{n-2} \\
\Delta A_{n+1} & \Delta A_{n}  & \Delta A_{n-1}
\end{array} \right \vert
}
\eea
This second order transformation returns the exacts when applying to the Taylor expansions of rational functions with two poles.

\section{Estimation of Error $\delta c_1$}
\label{app:error}

To estimate the error of $c_1$, we note that if we use the exact near field and far field expansions in Eq.~(\ref{matching-condition}), we would have found the exact value for the coefficient $c_1^{\rm ex}$, independent of the matching point $z^*$.
Mathematically we have:
\bea
\Psi_{\rm far}^{\rm ex}(z^*,c_1^{\rm ex})
= \Psi_{\rm near}^{\rm ex}(z^*),
\label{exact equation}
\eea
where the superscript ex stands for ``exact''. By contrast, because of the approximate nature of the expansions we use, the coefficient $c_1$ we find is different from its exact value.  Let $c_1 = c_1^{\rm ex} + \delta c_1$, where $\delta c_1$ is the error, we can rewrite Eq.~(\ref{matching-condition}) in the following form:
 \be
\Psi^{\rm app}_{\rm far}(z^*, c_1^{\rm ex} + \delta c_1)
= \Psi^{\rm app}_{\rm near}(z^*).
\ee
Expanding the left hand side to the linear order of $\delta c_1$, we have:
\bea
 \Psi^{\rm app}_{\rm far}(z^*, c_1^{\rm ex} ) +
 \delta c_1 \, \frac{ \partial \Psi^{\rm app}_{\rm far}(z^*, c_1^{\rm ex} )}{\partial c_1}
=  \Psi^{\rm app}_{\rm near}(z^*, c_1^{\rm ex}).
\nonumber\\
 \label{approximate equation}
\eea
The derivative of $\Psi^{\rm app}_{\rm far}$ with respect to $c_1$ can be easily calculated using Mathematica.   We further introduce the error of the far field and near field expansions (at the matching point $z^*$) as following:
\begin{subequations}
\label{delta-Psi}
\bea
\delta \Psi_{\rm far}(z^*,c_1) &=&
\Psi^{\rm app}_{\rm far}(z^*,c_1)
- \Psi_{\rm far}(z^*,c_1), \\
\delta \Psi_{\rm near}(z^*) &=&
\Psi^{\rm app}_{\rm near}(z^*)
- \Psi_{\rm near}(z^*).
\eea
\end{subequations}
These errors can be estimated by the difference between the approximation at the current order and that at the next order, which can be conveniently calculated using Mathematica.  Using Eqs.~(\ref{exact equation}, \ref{approximate equation}, \ref{delta-Psi}), we can estimate the upper bound of the error $\delta c_1$ as
\be
|\delta c_1| \leq
	\frac{|\delta \Psi^{\rm app}_{\rm near}(z^*)|
+ |\delta \Psi^{\rm app}_{\rm far}(z^*,c_1)|}
{ |\frac{\partial \Psi^{\rm app}_{\rm far}(z^*,c_1)}{\partial c_1}|}.  
\ee


\section{Coefficients of Near Field Expansions}

\begin{table*}
\vspace{-5mm}
\begin{sideways}
\begin{tabular}{cccccccccccccccccccccccc}
\hline\hline \noalign{\smallskip}
Electrolyte
& \quad  Plate & & & & Near Field &Coefficients & & & & & & & &
\\  \noalign{\smallskip}\hline \noalign{\smallskip}

\multirow{4}{*}
{$3:-1$} &
\multirow{2}{*}
 {$+$}
  &$a_2$ & $a_4$ & $a_6$ & $a_8$ & $a_{10}$ & $a_{12}$
& $a_{14}$ & $a_{16}$
& $a_{18}$
\\
&
&$5.55556\times10^{-2}$&$3.08642\times10^{-4}$ &$3.26605\times10^{-6}$
&$-9.00142\times10^{-6}$
&$9.13908\times10^{-7}$
&$-4.81582\times10^{-8}$
&$1.68532\times10^{-9}$&$-9.85359\times10^{-11}$
&$1.24603\times10^{-11}$
\\
& \multirow{2}{*}
{$-$}
&$a_2$ & $a_{8/3}$ & $a_{10/3}$ & $a_6$ & $a_{20/3}$ & $a_{22/3}$
& $a_{8}$ & $a_{26/3}$
& $a_{28/3}$
\\
&
&$-1.66667\times10^{-1}$&$7.37514\times10^{-2}$ &$0$
&$-8.33333\times10^{-3}$
&$2.89221\times10^{-3}$
&$-1.43139\times10^{-4}$
&$-7.93651\times10^{-4}$&$7.27245\times10^{-4}$
&$2.53720\times10^{-4}$
\\ \noalign{\smallskip} \hline \noalign{\smallskip}
\multirow{4}{*}
 {$4:-1$} &
\multirow{2}{*}
{$+$}
 &$a_2$ & $a_4$ & $a_6$ & $a_8$ & $a_{10}$ & $a_{12}$
& $a_{14}$ & $a_{16}$
& $a_{18}$
\\
&
&$4.16667\times10^{-2}$&$1.73611\times10^{-4}$ &$1.37787\times10^{-6}$
&$1.29175\times10^{-8}$
&$-2.27142\times10^{-7}$
&$2.54967\times10^{-8}$
&$-1.45686\times10^{-9}$&$5.52515\times10^{-11}$
&$-1.54541\times10^{-12}$
\\
& \multirow{2}{*}
{$-$}
&$a_2$ & $a_{5/2}$ & $a_{6}$ & $a_{13/2}$ & $a_{7}$ & $a_{15/2}$
& $a_{8}$ & $a_{17/2}$
& $a_{9}$
\\
&
&$-1.66667\times10^{-1}$&$9.08881\times10^{-2}$ &$0$&$0$
&$-1.11111\times10^{-2}$
&$6.88546\times10^{-3}$
&$-1.03258\times10^{-3}$
&$0$&$-1.41093\times10^{-3}$
\\ \noalign{\smallskip} \hline \noalign{\smallskip}
\multirow{4}{*}
 {$3:-2$} &
\multirow{2}{*}
{$+$}
 &$a_2$ & $a_4$ & $a_5$ & $a_6$ & $a_7$ & $a_8$
& $a_9$ & $a_{10}$
& $a_{11}$
\\
&
&$5.55556\times10^{-2}$&$6.17284\times10^{-4}$ &$-9.93808\times10^{-4}$&$1.30642\times10^{-4}$
&$6.90144\times10^{-5}$&$3.26605\times10^{-7}$
&$-3.06731\times10^{-6}$
&$-5.74816\times10^{-7}$
&$7.91171\times10^{-8}$
\\
& \multirow{2}{*}
{$-$}
&$a_2$ & $a_{10/3}$ & $a_{4}$ & $a_{14/3}$ & $a_{16/3}$ & $a_{6}$
& $a_{20/3}$ & $a_{22/3}$
& $a_{8}$
\\
&
&$-8.33333\times10^{-2}$&$1.55126\times10^{-2}$ &$2.08333\times10^{-3}$&$0$
&$-3.40188\times10^{-4}$
&$9.92063\times10^{-5}$
&$5.75442\times10^{-5}$
&$4.19565\times10^{-5}$&$-7.68980\times10^{-6}$
\\ \noalign{\smallskip} \hline \noalign{\smallskip}
\multirow{4}{*}
 {$4:-3$} &
\multirow{2}{*}
{$+$}
 &$a_2$ & $a_4$ & $a_{14/3}$ &$a_6$& $a_{20/3}$ & $a_{22/3}$ &$a_8$& $a_{26/3}$
& $a_{28/3}$
\\
&
&$4.16667\times10^{-2}$&$5.208334\times10^{-4}$ &$-1.21226\times10^{-3}$&$1.24008\times10^{-5}$
&$7.68644\times10^{-5}$&$0$
&$3.48772\times10^{-7}$
&$-3.41031\times10^{-6}$
&$-1.11404\times10^{-6}$
\\
& \multirow{2}{*}
{$-$}
&$a_2$ & $a_{7/2}$ & $a_{4}$ & $a_{5}$& $a_{6}$ & $a_{13/2}$
& $a_{7}$ & $a_{15/2}$
& $a_{8}$
\\
&
&$-5.55556\times10^{-2}$&$8.27079\times10^{-3}$ &$1.23457\times10^{-3}$&$0$
&$-5.22568\times10^{-5}$&$0$
&$2.77899\times10^{-5}$
&$2.02947\times10^{-5}$&$-2.61284\times10^{-6}$
\\ \noalign{\smallskip} \hline \noalign{\smallskip}
\end{tabular}
\label{Table:near-field-coefficient}
\end{sideways}
\end{table*}

\end{document}